\documentclass{article}

     \PassOptionsToPackage{numbers, compress}{natbib}

\usepackage[preprint]{neurips_2025}
\usepackage{amsmath}
\usepackage{amssymb}
\usepackage{bm}
\usepackage{graphicx}
\usepackage{enumitem}
\usepackage{makecell}
\usepackage{longtable}
\usepackage{colortbl}
\usepackage{float}
\usepackage{capt-of}
\usepackage{caption} 
\usepackage[utf8]{inputenc} 
\usepackage[T1]{fontenc}    
\usepackage{hyperref}       
\usepackage{url}            
\usepackage{booktabs}       
\usepackage{amsfonts}       
\usepackage{nicefrac}       
\usepackage{microtype}      
\usepackage{xcolor}         

\usepackage{graphicx}

\usepackage{adjustbox}

\usepackage{amsfonts}
\usepackage{multirow} 
\usepackage{siunitx} 

\usepackage{algorithm}
\usepackage{algorithmic}

\definecolor{surface}{HTML}{ecedf7}
\definecolor{blue}{HTML}{0669dc}
\definecolor{gray}{HTML}{727785}
\definecolor{green}{HTML}{227000}
\definecolor{red}{HTML}{9c48ba}
\definecolor{error}{HTML}{ba1a1a}

\title{FinLoRA: Benchmarking LoRA Methods for Fine-Tuning LLMs on Financial Datasets}

\author{%
Dannong Wang$^{1}$ \quad Jaisal Patel$^{1}$ \quad Daochen Zha$^2$ \quad Steve Y. Yang$^3$ \quad Xiao-Yang Liu$^2$ \\
$^1$Rensselaer Polytechnic Institute \quad 
$^2$Columbia University \quad
$^3$Stevens Institute of Technology  \\
Emails: \texttt{\{wangd12, patelj8\}@rpi.edu}, 
\texttt{syang14@stevens.edu}, 
\texttt{xl2427@columbia.edu}
}

\begin{document}

\maketitle

\begin{abstract}

Low-rank adaptation (LoRA) methods show great potential for scaling pre-trained general-purpose Large Language Models (LLMs) to hundreds or thousands of use scenarios. However, their efficacy in high-stakes domains like finance is rarely explored, e.g., passing CFA exams and analyzing SEC filings. In this paper, we present the open-source FinLoRA project that benchmarks LoRA methods on both general and highly professional financial tasks. First, we curated 19 datasets covering diverse financial applications; in particular, we created four novel XBRL analysis datasets based on 150 SEC filings. Second, we evaluated five LoRA methods and five base LLMs. Finally, we provide extensive experimental results in terms of accuracy, F1, and BERTScore and report computational cost in terms of time and GPU memory during fine-tuning and inference stages. We find that LoRA methods achieved substantial performance gains of 36\% on average over base models. Our FinLoRA project provides an affordable and scalable approach to democratize financial intelligence to the general public. Datasets, LoRA adapters, code, and documentation are available at \url{https://github.com/Open-Finance-Lab/FinLoRA} 

\end{abstract}

\section{Introduction}


Large language models (LLMs)~\cite{zhao2023survey,zheng2023judging} have demonstrated impressive general capabilities in various vertical domains, such as finance~\cite{wu2023bloomberggpt,liu2023data,Lee_2025,bhatia-etal-2024-fintral}, healthcare~\cite{wang2024twin,lu2024uncertainty,zhou2024large}, law~\cite{DBLP:conf/acl/WuWGL24}, education~\cite{liu2024socraticlm}, and scientific discovery~\cite{lu2022cot,chen2021data}. In the finance sector, LLMs have been applied to tasks such as sentiment analysis~\cite{zhang2023enhancingfinancialsentimentanalysis}, question-response, and stock market prediction~\cite{Lee_2025}.


Cost-effective adaptation is critical for applying LLMs to vertical domains like finance, since general-purpose LLMs lack the specialized knowledge to excel in professional-level tasks.  Full fine-tuning can close such performance gaps but is prohibitive for most organizations due to its computationally demanding nature. As such, parameter-efficient fine-tuning (PEFT), particularly Low-Rank Adaptation (LoRA) \cite{lora} and its variants~\cite{qlora,Mao_2024,yang2024lowrankadaptationfoundationmodels,huanlora,liu2022few,chen2022adaptformer,meng2024pissa}, has emerged as an affordable and scalable solution. LoRA methods can enhance pre-trained general-purpose LLMs with domain-specific knowledge and improve performance on downstream tasks~\cite{Mao_2024}. 


Recent research like FinGPT~\cite{liu2023data,icdcs} has applied a quantized LoRA method \cite{qlora} to general financial tasks; however,
the comparative performance of various LoRA variants in complex, professional-level financial tasks remains rarely explored. Previous research shows that LLMs are struggling with professional-level financial tasks, such as analyzing SEC filings ~\cite{islam2023financebenchnewbenchmarkfinancial} and passing financial certificate exams \cite{callanan-etal-2024-gpt}. A critical area within professional finance involves eXtensible Business Reporting Language (XBRL) data~\cite{xbrl}, the de facto global standard for business reporting. Despite XBRL's importance, dedicated datasets for related analytical tasks are scarce. This deficiency, coupled with the need to evaluate different LoRA methods on highly specialized financial tasks, motivates our introduction of FinLoRA: a comprehensive benchmark designed to assess LoRA variants across diverse financial scenarios, with an emphasis on professional XBRL applications.

\begin{figure}[t]
    \centering
    \includegraphics[width=16cm]{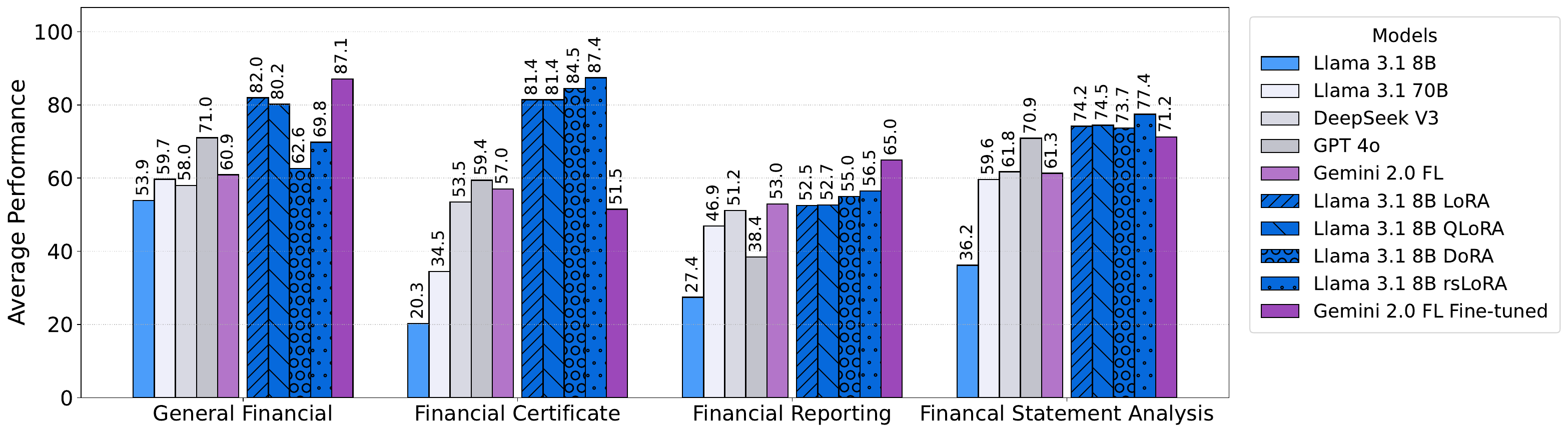}
    \caption{Average performance of base models and LoRA models.}
    \label{fig:overall}
\end{figure}

This paper demonstrates that fine-tuning state-of-the-art LLMs can significantly improve performance across a range of financial tasks, including specialized XBRL analysis, locally and cost-effectively using widely accessible GPUs. As illustrated in Fig.~\ref{fig:overall}, our LoRA-adapted models achieve notable performance improvements over baseline models across four categories of financial tasks. Our main contributions are summarized as follows:
\begin{itemize}[leftmargin=*]
    \item We curated 19 financial datasets, including general financial tasks, financial analysis, and professional-level XBRL tasks. In particular, we created four novel XBRL analysis datasets. This enables future research to perform rigorous evaluation of LoRA methods in financial tasks.
    \item We implemented and fairly compared five LoRA methods—including LoRA~\cite{lora}, QLoRA~\cite{qlora}, DoRA~\cite{liu2024dora}, rsLoRA~\cite{kalajdzievski2023rankstabilizationscalingfactor}, and Federated LoRA—by fine-tuning models on financial datasets. LoRA methods achieved  an average increase of 36\% in accuracy over baseline models, which validates the effectiveness of low-rank adaptation and quantization for fine-tuning LLMs.
    \item We conducted an extensive analysis with 46 rounds of fine-tuning and 194 rounds of evaluations for LoRA methods from four angles: (i) a comprehensive comparison across different base models and datasets, (ii) performance on various types of financial tasks, (iii) resource requirements for fine-tuning and inference, and (iv) practical considerations for LoRA deployment in finance.
\end{itemize}

Our benchmark is open-sourced at \url{https://github.com/Open-Finance-Lab/FinLoRA}.


\section{Is Fine-tuning of LLMs Needed on Financial Tasks?}

\label{section:preliminaries}





While general-purpose LLMs—such as GPT-4o~\cite{hurst2024gpt}, Llama 3.1~\cite{llama}, and DeepSeek-V3~\cite{liu2024deepseek}—demonstrate broad NLP competence, their performance often falls short on nuanced financial tasks ~\cite{islam2023financebenchnewbenchmarkfinancial, callanan-etal-2024-gpt}. This section discusses three key reasons that underscore the necessity of fine-tuning, particularly with methods like LoRA, for developing effective financial LLMs: 
\textbf{\textit{(i)} Lack of High-Quality Financial Data in Pre-Training Datasets.} 
Many pre-training datasets, such as The Pile~\cite{gao2020pile}, primarily draw from general web crawls (e.g., GitHub, arXiv). These sources often under-represent high-quality, specialized financial data, which may be private and exist in complex formats (like XBRL). Consequently, to equip LLMs with the understanding required for complex financial analysis, targeted fine-tuning on curated, domain-specific datasets becomes essential.
\textbf{(\textit{ii)} General LLMs' Failure in Specialized Financial Tasks.}
General LLMs often struggle with specialized tasks that demand deep domain-specific knowledge. XBRL analysis provides a clear illustration of these difficulties: 
Table \ref{tab:case-terminology} details the typical Llama base model's errors on XBRL questions and its improved outputs after LoRA fine-tuning. In Question 1, the base model mistags a \$2.0 billion value because it relies on superficial keyword matches (e.g., "equity," "carrying value,") and applies the generic tag \textit{us-gaap:MajorityEquityInterest}, ignoring the context "under the equity method." 
In Question 2, the base model incorrectly selects a tag referencing the value 1,209,000,000 by matching the keyword "Equity" and also ignores the decimals="-6" attribute. \textbf{\textit{(iii)}~Cost and Time.} As shown in Table \ref{cost}, the training-from-scratch approach of BloombergGPT~\cite{wu2023bloomberggpt}, which reportedly cost \$2.7 million and took 53 days to train (Table \ref{cost}), is economically nonviable for most organizations. In contrast, fine-tuning existing foundational models using LoRA methods is significantly more accessible and time-efficient.

\begin{table}[t]
\small
\caption{Case study—XBRL tagging (Google 10-Q 2025-Q1) and XBRL formula calculation (Travelers 10-K FY-2023)—from base Llama 3.1 8B Instruct and our LoRA fine-tuned version.}
\label{tab:case-terminology}
\centering

\begin{tabular}{p{0.28\linewidth}|p{0.65\linewidth}}
\toprule \rowcolor{surface}
\textbf{Question 1} & 
What is the appropriate XBRL US-GAAP tag for “2.0” in  
\textbf{“...equity securities accounted for under the equity method had a carrying value of approximately \$2.0 billion”}\,?  \\ \midrule
{Llama 3.1 8B } & \texttt{us-gaap:\textcolor{error}{MajorityEquityInterest}} \\ \midrule \rowcolor{surface}
{Llama 3.1 8B LoRA (8bit r8)} & \texttt{us-gaap:EquityMethodInvestments} \\ \midrule
{Ground truth} & \texttt{us-gaap:EquityMethodInvestments} \\ \midrule \rowcolor{surface}
\textbf{Question 2} &
What is Travelers Companies Inc’s \textbf{Equity Multiplier} for FY 2023?  
(Answer with a formula substituted with values.) \{XBRL Context\} \\ \midrule

{Llama 3.1 8B} &
\(\textcolor{error}{(1{,}209{,}000{,}000} \;/\; 249{,}210{,}000{,}000\textcolor{error}{)}\) \\ \midrule \rowcolor{surface}

{Llama 3.1 8B LoRA (8bit r8)} &
\(125,978,000,000 \;/\; 249,210,000,000\) \\ \midrule

{Ground truth} &
\(125,978,000,000 \;/\; 249,210,000,000\) \\ \bottomrule
\end{tabular}
\end{table}

\section{FinLoRA Benchmark}
\label{section:benchmark}

\subsection{Benchmark Tasks, Datasets, and Metrics}

As displayed in Table~\ref{tasks}, we consider four types of tasks: general financial tasks, financial certificate, financial reporting, and financial statement analysis.

\paragraph{Public Financial Datasets}
FinLoRA includes 15 public financial datasets. 
\textit{(i)} Sentiment analysis (SA): Financial Phrase Bank (FPB)~\cite{fpb}, Financial QA Sentiment Analysis (FiQA SA)~\cite{fiqa}, Twitter Financial News Sentiment (TFNS)~\cite{tfns}, and News with GPT Instruction (NWGI)~\cite{icdcs}, each with financial text from news or tweets and sentiment labels. 
\textit{(ii)} Headline analysis: The Headline dataset~\cite{headline} classifies financial headlines based on various questions into two classes: "yes" and "no". 
\textit{(iii)} Named-entity recognition (NER): NER dataset~\cite{ner} annotates one entity per sentence, categorized into one of three classes: "location", "person", and "organization". 
\textit{(iv)} Financial certificate: CFA Level I, II, and III, and CPA Regulation. 
\textit{(v)} Financial reporting: XBRL Terminology~\cite{xbrlagent}, Financial Numeric Entity Recognition (FiNER)~\cite{loukas-etal-2022-finer}, and Financial Numeric Extreme Labeling (FNXL)~\cite{fnxl}. 
\textit{(vi)}: Financial statement analysis: Financial Math \cite{xbrlagent} and FinanceBench \cite{islam2023financebenchnewbenchmarkfinancial, xbrlagent}.

\paragraph{Newly-added XBRL Analysis Datasets}
We introduce 4 novel XBRL analysis datasets, i.e., extracting and analyzing SEC financial reports in XBRL format. These question-answering datasets, derived from the 2019-2023 annual reports of Dow Jones 30 companies, provide each example with a question, a relevant filtered XBRL text segment as source material, and a ground truth answer. The datasets cover four distinct task types:
\textit{(i)} \textbf{XBRL tag extraction} involves extracting a specific XBRL tag from a raw XBRL text segment given a natural language description of the tag.
 \textit{(ii)} \textbf{XBRL value extraction} focuses on extracting a numeric value from the raw XBRL text segment given a natural language description of the value. 
\textit{(iii)} \textbf{XBRL formula construction} tasks the LLM to first identify and select multiple relevant facts (and their corresponding XBRL tags) from the XBRL data, and then construct a standard financial formula (e.g., Net Profit Margin, Quick Ratio) using these selected tags as components.
\textit{(iv)} \textbf{XBRL formula calculation} builds on the previous task and requires the LLM to substitute the actual numeric values into the formula and compute the final result.

\paragraph{Dataset Construction Pipeline}
Initially, we classified financial tasks into nine categories, creating a training set for each to develop category-specific LoRA adapters per configuration. The four novel XBRL analysis datasets were constructed using XBRL-formatted 10-K annual reports from Dow Jones 30 companies (2019-2023). For these, we generated the four aforementioned types of questions by applying five distinct templates to consolidated, company-specific facts. To ensure contextual relevance, XBRL file segments were automatically filtered based on pertinent factors like year and reporting axes. Further details on the XBRL dataset creation and the processing of other public datasets are available in Appendix A.

\paragraph{Metrics} For all general financial tasks, financial analysis, XBRL tagging, financial math, and XBRL analysis tasks, we use Exact Match (EM) to evaluate the LLMs' output and report both the accuracy and weighted F1 score (in the supplementary materials). For XBRL Term and FinanceBench, we report BERTScore F1~\cite{zhang2020bertscoreevaluatingtextgeneration} instead. We also report the average of scores across the tasks with BERTScore F1 multiplied by 100.

\newcommand{\hf}[1]{\href{https://huggingface.co/datasets/#1}{HF}}
\newcommand{\github}[1]{\href{https://github.com/#1}{GitHub}}
\begin{table*}[]
\small
\caption{Benchmark tasks and datasets.\\}
\centering
\label{tasks}
\begin{adjustbox}{width=\textwidth}

\begin{tabular}{l|cccccc}
\toprule
\bf Datasets &  \bf Types & \bf \#Train/\#Test & \bf \shortstack{Average \\ Prompt \\ Length} & \bf Metrics & \bf \shortstack{ Sources \\ \& License} \\

\midrule 
\rowcolor{surface}
\multicolumn{6}{c}{\textbf{General Financial Tasks} (Total: 122.9k/31.7k)}   \\
\midrule
FPB~\cite{fpb} & Sentiment Analysis & 3.1k/970 & 56 & Accuracy, F1 & \hf{TheFinAI/en-fpb}, \tiny{CC BY-SA 3.0} \\
FiQA SA~\cite{fiqa}& Sentiment Analysis & 822/234 & 48 & Accuracy, F1 & \hf{TheFinAI/fiqa-sentiment-classification} \tiny{MIT} \\
TFNS~\cite{tfns}& Sentiment Analysis & 9.5k/2.4k & 52 & Accuracy, F1 & \hf{zeroshot/twitter-financial-news-sentiment} \tiny{MIT} \\
NWGI~\cite{liu2023data} & Sentiment Analysis & 12.9k/4.1k & 81 & Accuracy, F1 & \hf{TheFinAI/NWGI\_test} \tiny{MIT}\\
Headline~\cite{headline} & Headline Analysis & 82.2k/20.5k & 43 & Accuracy, F1 & \hf{FinGPT/fingpt-headline-cls} \tiny{ CC BY-SA 3.0}\\
NER~\cite{ner} & \shortstack{NER} & 13.5k/3.5k & 138 & Accuracy, F1 & \hf{FinGPT/fingpt-ner-cls} \tiny{ CC BY-SA 3.0}\\

\midrule 
\rowcolor{surface}
\multicolumn{6}{c}{\textbf{Financial Certificate Tasks} (Total: 472/346)}   \\
\midrule

CFA Level I & \shortstack{ Analyst Exam} & 180/90 & 181 & Accuracy, F1 & \multirow{4}*{\shortstack{Internet \\\tiny{(Public; Not} \\ \tiny{Released Due to} \\ \tiny{Copyright})}} \\
CFA Level II & \shortstack{ Analyst Exam} & 88/77 & 1.0k & Accuracy, F1 &  \\
CFA Level III & \shortstack{ Analyst Exam} & 80/78 & 961 & Accuracy, F1 &  \\
CPA REG & \shortstack{Accountant Exam} & 124/101 & 147 & Accuracy, F1 &  \\

\midrule 
\rowcolor{surface}
\multicolumn{6}{c}{\textbf{Financial Reporting Tasks} (Total: 15.9k/8.3k)} \\
\midrule
FiNER-139~\cite{loukas-etal-2022-finer} & XBRL Tagging & 10.0k/7.4k & 1.8k & Accuracy, F1 & \hf{nlpaueb/finer-139}  \tiny{ CC BY-SA 4.0}\\
FNXL~\cite{fnxl} & XBRL Tagging & -/247 & 7.1k & Accuracy, F1 & \github{soummyaah/FNXL}  \tiny{ Public}\\
XBRL Term~\cite{xbrlagent} & Terminology & 5.9k/651 & 25 & BERTScore & \github{KirkHan0920/XBRL-Agent/blob/main/Datasets/XBRL\%20Terminology.xlsx} \tiny{MIT}\\
\midrule
\rowcolor{surface}
\multicolumn{6}{c}{\textbf{Financial Statement Analysis Tasks} (Total: 27.9k/7.3k)} \\
\midrule
Financial Math~\cite{xbrlagent} & Math & 800/200 & 116 & Accuracy & \github{KirkHan0920/XBRL-Agent/blob/main/Datasets/formulas\_with\_explanations\_with\_questions\_with\_gt.xlsx} \tiny{MIT} \\
FinanceBench~\cite{islam2023financebenchnewbenchmarkfinancial,xbrlagent} & Math & 86/43 & 983 & BERTScore & \github{KirkHan0920/XBRL-Agent/blob/main/Datasets/financebench.xlsx} \tiny{CC BY-NC 4.0} \\
Tags Extraction & XBRL Analysis & 10.1K/2.9k & 3.8k & Accuracy, F1 & \hf{wangd12/XBRL_analysis} \tiny{MIT}\\
Values Extraction & XBRL Analysis & 10.1k/2.5k & 3.8k & Accuracy, F1 & \hf{wangd12/XBRL_analysis} \tiny{MIT}\\
Formula Construction & XBRL Analysis & 3.4K/835 & 3.8k & Accuracy, F1 & \hf{wangd12/XBRL_analysis} \tiny{MIT}\\
Formula Calculation & XBRL Analysis & 3.4K/835 & 3.8k & Accuracy, F1 & \hf{wangd12/XBRL_analysis} \tiny{MIT}\\

\bottomrule
\end{tabular}
\end{adjustbox}
\end{table*}

\subsection{Base Models and LoRA Methods}

\paragraph{Base Models}
We benchmark two models for both base model and LoRA fine-tuning performance—Llama 3.1 8B Instruct~\cite{llama} and Gemini 2.0 Flash Lite~\cite{geminiteam2024geminifamilyhighlycapable}. We also evaluated three additional models—Llama 3.1 70B Instruct~\cite{llama}, DeepSeek V3~\cite{liu2024deepseek}, and GPT-4o~\cite{hurst2024gpt}—as base models only. 

\paragraph{LoRA Methods} We considered the following five popular LoRA methods.
\begin{itemize}[leftmargin=*]
    \item (Vanilla) \textbf{LoRA}: Low-rank adaptation (LoRA)~\cite{lora} is a parameter-efficient fine-tuning method that preserves the weights of the pre-trained model and introduces a smaller set of trainable weights. The updated weights follow the low-rank decompositions $\Delta \bm{W} = \gamma_r\bm{B}\bm{A}$, where $\gamma_r$ is a scaling factor ($\gamma_r=\frac{\alpha}{r}$ with  $\alpha$ > 0 and rank $r$ > 0), $\bm{A} \in \mathbb{R}^{r \times k}$ and $\bm{B} \in \mathbb{R}^{d \times r}$ are trainable parameters, and $\bm{W}_0 \in \mathbb{R}^{d \times k}$ denote the pre-trained weights. 
    During the fine-tuning stage, the forward pass is $\bm y=\bigl(\bm W_{0}\;+\;\gamma_{r}\bm B\bm A\bigr)\bm x=\bm W_{0}\bm x\;+\;\gamma_{r}\bm B\bm A\bm x$. 
    
    \item \textbf{QLoRA}. Quantized LoRA (QLoRA)~\cite{qlora} further reduces memory usage by using 4-bit quantization. During fine-tuning, all weights of the pre-trained model are quantized to 4 bits. Weights will be dynamically dequantized back to 16 bits when performing computation with the input sequence $\bm{x}$ and the adapter matrix $\bm{A}$ and $\bm{B}$, which remain in $16$-bit precision throughout the process, where
    $\bm{y} = p_{16}(\bm{W}_0^{\text{NF4}}) \bm{x} + \gamma_r \bm{B} \bm{A} \bm{x}$.
    The process is similar in the inference stage, where the merged weights $\bm{W}$ are loaded in $4$-bit precision. 
    \item \textbf{DoRA}. Weight-Decomposed Low-Rank Adaptation (DoRA) \cite{liu2024dora} decomposes $\bm{W}_0 \in \mathbb{R}^{d\times k}$ into a column-wise magnitude vector $\bm m\in\mathbb R^{1\times k}$ and a direction matrix $\bm V \in \mathbb{R}^{d \times k}$, where $\bm{m}=\|\bm{W}_0\|_c$ (with $\| \cdot \|$ being column-wise norm) and $\bm{V}=\bm{W}_0$. Only the direction matrix receives updates through LoRA. The magnitude vector is updated separately.
    DoRA can achieve accuracy close to that from full fine-tuning while keeping the same parameter count as LoRA. 
    \item \textbf{rsLoRA}. Vanilla LoRA uses a scaling factor $\nicefrac{\alpha}{r}$, which may cause gradients to explode or diminish as the rank $r$ increases. Rank-Stabilized LoRA (rsLoRA) \cite{kalajdzievski2023rankstabilizationscalingfactor} uses a scaling factor $\nicefrac{\alpha}{\sqrt{r}}$:
    $       \bm W'=\bm W_0+\frac{\alpha}{\sqrt{r}}\bm B\bm A.
    $
    This scaling results in gradient-scale stability at higher ranks, enabling the rank to be higher for long-context tasks like XBRL analysis. 
    \item \textbf{LoRA with Federated Learning}. In the finance sector, multiple institutions may want to collaborate using their own proprietary datasets, but they cannot share their data due to compliance reasons and privacy concerns. Federated learning solves this issue by fine-tuning a model on local data and aggregating LoRA updates to a central node. 
\end{itemize}

\subsection{Benchmark Angles}

\paragraph{Angle I: LoRA Methods' Performance on Financial Datasets}
We seek to learn which LoRA method is most effective in financial tasks, in terms of both category-specific and overall performance, and how these LoRA fine-tuned models perform compared to existing state-of-the-art (SOTA) models. We fine-tuned Llama 3.1 8B Instruct using LoRA, QLoRA, rsLoRA, and DoRA, representing open-source models and fine-tuning approaches, and fine-tuned Gemini 2.0 Flash Lite using Google's proprietary fine-tuning methods as a baseline representing closed-source counterparts. 

\paragraph{Angle II: LoRA Suitability for Financial Tasks}
We wish to investigate how the benefits of LoRA fine-tuning vary across different financial tasks. This angle is motivated by the need to identify which specific applications (e.g., sentiment analysis, XBRL tagging, XBRL analysis) are most responsive to fine-tuning, and what properties of the datasets cause this. 

\paragraph{Angle III: Resources of LoRA Fine-tuning and Inference}
We aim to compare which LoRA methods, out of the tested methods, are the most cost-effective in fine-tuning and compare the fine-tuning cost to closed-source fine-tuning services. 
We are also motivated to measure and compare the inference speeds of LoRA-fine-tuned models against their larger base model counterparts. The goal is to quantify the potential for reduced latency and increased throughput, which are critical for real-time financial applications and operational efficiency.

\paragraph{Angle IV: Practical Considerations for LoRA Deployment in Finance}
To assess the viability of deploying LoRA-fine-tuned models in real-world financial scenarios, we investigate two key concerns: \textit{(i)} Data Privacy in Collaborative Training: While local LoRA fine-tuning enhances data protection, collaborative model training across multiple institutions often requires approaches like Federated Learning to preserve the privacy of proprietary training data. We investigate this by simulating data distribution across several nodes and evaluating LoRA fine-tuning performance against centralized training. \textit{(ii)} Catastrophic Forgetting: Fine-tuning can risk degrading a model's pre-existing general knowledge and capabilities. To quantify this, we evaluate our LoRA-fine-tuned models on established general-domain benchmarks, such as MMLU \cite{hendrycks2020measuring}, measuring any performance changes on tasks outside their financial fine-tuning scope.

\newcommand{\acc}{\textcolor{blue}} 
\newcommand{\fscore}{\textcolor{gray}} 
\newcommand{\bscore}{\textcolor{orange}} 
\newcommand{\precision}{\textcolor{red}} 

\begin{table*}[H]
\small
\centering
\caption{Fine-tuning LLMs with LoRA methods: the number of parameters and GPU memory.}
\begin{tabular}{l|ccc|cc}
\toprule
%

Models   & Parameters  & \shortstack{GPU Memory \\ Batch = 4} &  \shortstack{GPU Memory \\ Batch = 8} & Model size & Percentage\\
\midrule
Llama-3.1-8B-16bit (base) & 8.03 B  & - & - & 16.06 GB  & - \\

\\[-2.2ex]\hline\\[-1.9ex]

Llama-3.1-8B-r8-16bit & 4.72 M  & 30.91 GB & 30.91 GB& 16.08 GB & 100.1\%\\
Llama-3.1-8B-r8-8bit & 4.72 M  & 11.41 GB&  11.81 GB& 8.04 GB & 50.1\% \\
Llama-3.1-8B-r8-4bit & 4.72 M  & 8.26 GB& 8.65 GB& 4.02 GB & 25.0\% \\

\\[-2.2ex]\hline\\[-1.9ex]

Llama-3.1-8B-r4-16bit & 2.36 M  & 30.90 GB& 30.90 GB& 16.07 GB & 100.1\% \\
Llama-3.1-8B-r4-8bit & 2.36 M  & 11.40 GB& 11.78 GB& 8.03 GB & 50.0\%\\
Llama-3.1-8B-r4-4bit & 2.36 M  & 8.25 GB& 8.61 GB& 4.02 GB  & 25.0\% \\





\bottomrule
\end{tabular}
\end{table*}

\section{Benchmark Results}

\label{section:results}

%

\paragraph{Setup} Our experiments were conducted on four NVIDIA A5000 GPUs. For closed-source models, we employed various inference and fine-tuning APIs. For each LoRA method, we fine-tuned 9 LoRA adapters based on their respective training sets merged by task categories. We used a learning rate of 1e-4 and a batch size of 2--8 based on prompt length (Refer to Appendix C for details).  For inference, we used a temperature of 0.0. Overall, we conducted 46 rounds of fine-tuning and 194 rounds of evaluations to benchmark these LoRA methods from different angles.

\subsection{Angle I: LoRA Methods Performance on Financial Datasets} \label{section:overall}

\paragraph{Comparative Performance of LoRA Variants} Table \ref{tab:main_results_consolidated} shows the performance of base models and different LoRA fine-tuned models. Vanilla LoRA (8-bit, rank 8) achieves the highest overall average score (74.74), a 37.69\% increase over the Llama 3.1 8B base model's 37.05. Fig.~\ref{fig:overall} shows the performance by category. Vanilla LoRA outperforms other LoRA variants in general financial tasks, while rsLoRA leads in financial analysis, financial reporting, and financial statement analysis.  

\paragraph{rsLoRA Performs Better at High Ranks} rsLoRA scales with \(\alpha/\sqrt{r}\) instead of \(\alpha/r\) to prevent gradient exploding or vanishing at large ranks. We set $r=8$ for memory efficiency. rsLoRA just slightly underperforms against LoRA and QLoRA. The rsLoRA paper's experiments \cite{kalajdzievski2023rankstabilizationscalingfactor} led to lower perplexity at higher ranks (e.g., $r$ = 64). This lower perplexity and the fact that higher rank LoRA captures more details suggest rsLoRA's benefits are primarily exploited at high ranks.

\paragraph{DoRA Benefits from Two Learning Rates} DoRA performed worse than the other three LoRA methods. We used the same learning rate for updating the magnitude vector and direction matrix. However, as shown in Table \ref{tab:main_results_consolidated}, this can lead to sub-optimal performance in some cases due to the gradient scales being different between the two types of updates in DoRA. This leads to DoRA sometimes under-training the magnitude vector in our experiments, which uses the same low learning rate. Thus, DoRA may achieve higher performance if the magnitude vector has its own learning rate that is higher than the low-rank update's learning rate.


\paragraph{LoRA-Tuned Llama 3.1 8B vs. Baseline Models and Gemini Fine-Tuned }
Compared to SOTA base LLMs, the LoRA-tuned Llama 3.1 8B Instruct models generally show superior performance across most datasets, with NWGI and FNXL being the exceptions. Against another fine-tuned baseline, the Gemini 2.0 FL fine-tuned model, this Gemini model excels in general financial tasks and XBRL data reporting. However, our Llama 3.1 8B Instruct LoRA variants demonstrate stronger average performance in financial analysis and XBRL data analysis tasks.

\subsection{Angle II: Financial Task LoRA Suitability} \label{section:suit}

\begin{figure}[t]
    \centering
    \includegraphics[width=14cm]{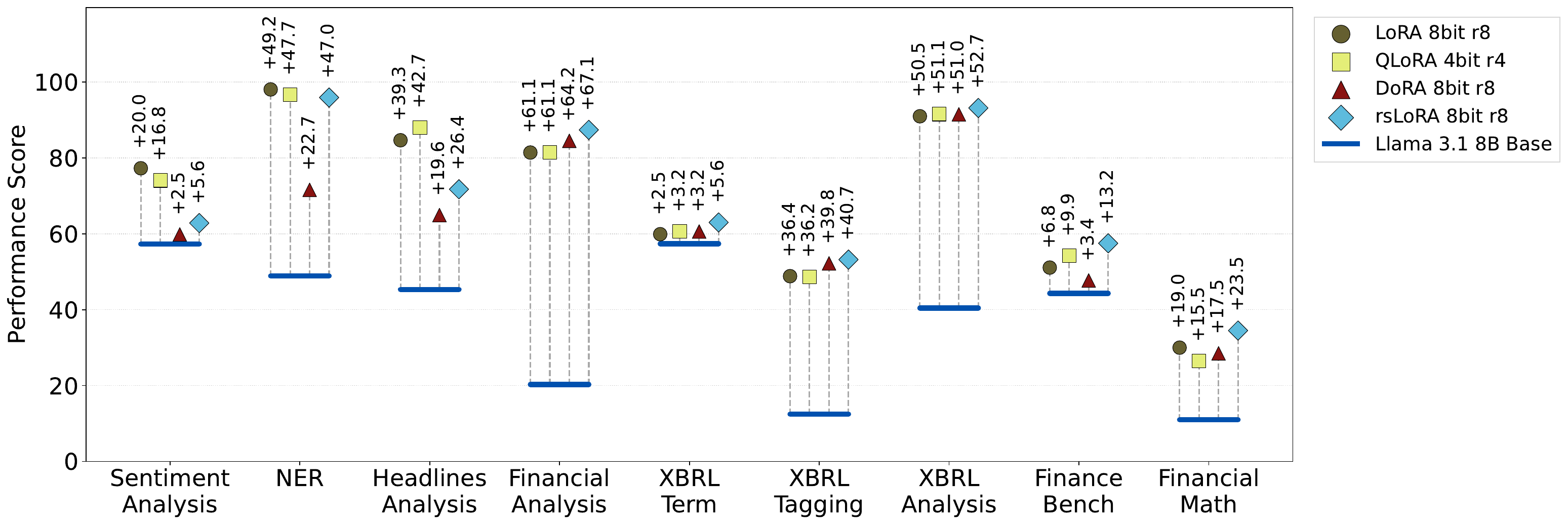}
    \caption{Task suitability.}
    \label{fig:suitability}
\end{figure}

Fig.~\ref{fig:suitability} highlights LoRA's varying effectiveness across different financial tasks. A key observation is the contrast in LoRA method improvements between XBRL Analysis tasks and FinanceBench. Although both aim to analyze financial statements, tasks based on XBRL data demonstrate substantial LoRA-induced performance improvements, whereas FinanceBench exhibits minimal gains. This disparity underscores XBRL's superior suitability for financial statement analysis. The standardized semantics and taxonomy inherent in XBRL likely provide a more structured and consistent learning environment for LLMs, facilitating more effective adaptation compared to FinanceBench, which relies on OCR-processed PDF data lacking such rich, standardized metadata. These findings emphasize the crucial role of XBRL in enabling effective LLM integration for financial report analysis.

\newcommand{\af}[4]{%
  \makecell{
    \textcolor{blue}{#1}\\          
    \textcolor{gray}{\scriptsize#3} \\          
  }%
}

\newcommand{\bert}[4]{%
  \makecell{
    \textcolor{green}{#3} \\            
  }%
}
\addtolength{\tabcolsep}{-0.2em}

\clearpage

\begin{table}[H] \small
\centering
\caption{Performance on financial tasks: \textcolor{blue}{accuracy} in blue, \textcolor{gray}{F1} in gray, and \textcolor{green}{BERTScore F1} in green. }
\begin{tabular}{l|ccccc|cccc|c}

\toprule
\multirow{5}{*}{\bf Datasets} & \multicolumn{5}{c|}{\bf Base Models} & \multicolumn{5}{c}{\bf Fine-tuned Models} \\
\cmidrule(lr){2-11}
 & Llama 3.1  & Llama 3.1 & DeepSeek & GPT- & Gemini 2.0 & Llama 3.1 & Llama 3.1  & Llama 3.1 & Llama 3.1 & Gemini \\
 & 8B \cite{llama} & 70B  \cite{llama} & V3 \cite{liu2024deepseek} & 4o \cite{hurst2024gpt} & FL \cite{geminiteam2024geminifamilyhighlycapable} & 8B  & 8B  & 8B  & 8B  & 2.0 FL \\ 
 &     &   &   &   &   & LoRA & QLoRA & DoRA & rsLoRA & N/A \\ 
 &     &   &   &   &   & 8bit-r8 & 4bit-r4 & 8bit-r8 & 8bit-r8 & N/A \\

\midrule

\rowcolor{surface}
\multicolumn{11}{c}{\bf General Financial Tasks} \\
\midrule
FPB 
& \af{68.73}{0.00}{0.677}{0.00} 
& \af{74.50}{0.00}{0.736}{0.00} 
& \af{78.76}{0.00}{0.764}{0.00} 
& \af{81.13}{0.00}{0.818}{0.00} 
& \af{81.02}{0.00}{0.894}{0.00} 
& \af{85.64}{0.00}{\textbf{0.922}}{0.00}
& \af{84.16}{0.00}{0.909}{0.00} 
& \af{81.93}{0.00}{0.901}{0.00} 
& \af{82.84}{0.00}{0.853}{0.00} 
& \af{\textbf{87.62}}{0.00}{0.878}{0.00} \\ 

\makecell[l]{FiQA SA}
& \af{46.55}{0.00}{0.557}{0.00} 
& \af{47.27}{0.00}{0.565}{0.00} 
& \af{60.43}{0.00}{0.686}{0.00} 
& \af{72.34}{0.00}{0.773}{0.00} 
& \af{68.09}{0.00}{0.810}{0.00} 
& \af{81.28}{0.00}{\textbf{0.884}}{0.00} 
& \af{78.30}{0.00}{0.874}{0.00}
& \af{78.72}{0.00}{0.874}{0.00} 
& \af{73.19}{0.00}{0.806}{0.00} 
& \af{\textbf{88.09}}{0.00}{0.879}{0.00} \\ 

TFNS 
& \af{69.97}{0.00}{0.683}{0.00} 
& \af{68.42}{0.00}{0.686}{0.00} 
& \af{84.38}{0.00}{0.846}{0.00} 
& \af{73.32}{0.00}{0.740}{0.00} 
& \af{26.38}{0.00}{0.385}{0.00} 
& \af{88.02}{0.00}{\textbf{0.932}}{0.00} 
& \af{83.84}{0.00}{0.910}{0.00}
& \af{59.09}{0.00}{0.702}{0.00} 
& \af{59.51}{0.00}{0.655}{0.00} 
& \af{\textbf{89.49}}{0.00}{0.896}{0.00} \\ 

NWGI 
& \af{43.86}{0.00}{0.583}{0.00} 
& \af{50.14}{0.00}{0.596}{0.00} 
& \af{7.44}{0.00}{0.097}{0.00} 
& \af{66.61}{0.00}{0.656}{0.00} 
& \af{48.16}{0.00}{0.614}{0.00} 
& \af{54.16}{0.00}{\textbf{0.690}}{0.00} 
& \af{49.96}{0.00}{0.645}{0.00}
& \af{19.57}{0.00}{0.281}{0.00} 
& \af{35.80}{0.00}{0.464}{0.00} 
& \af{\textbf{62.59}}{0.00}{0.581}{0.00} \\ \addlinespace

NER 
& \af{48.89}{0.00}{0.569}{0.00} 
& \af{46.28}{0.00}{0.454}{0.00} 
& \af{40.82}{0.00}{0.360}{0.00} 
& \af{52.11}{0.00}{0.523}{0.00} 
& \af{65.13}{0.00}{0.769}{0.00} 
& \af{\textbf{98.05}}{0.00}{\textbf{0.981}}{0.00} 
& \af{96.63}{0.00}{0.966}{0.00}
& \af{71.59}{0.00}{0.834}{0.00} 
& \af{95.92}{0.00}{0.963}{0.00} 
& \af{97.29}{0.00}{0.973}{0.00} \\ \addlinespace

Headline 
& \af{45.34}{0.00}{0.558}{0.00} 
& \af{71.68}{0.00}{0.729}{0.00} 
& \af{76.06}{0.00}{0.779}{0.00} 
& \af{80.53}{0.00}{0.814}{0.00} 
& \af{76.60}{0.00}{0.847}{0.00} 
& \af{84.66}{0.00}{0.852}{0.00} 
& \af{88.03}{0.00}{0.886}{0.00}
& \af{64.93}{0.00}{0.781}{0.00} 
& \af{71.75}{0.00}{0.828}{0.00} 
& \af{\textbf{97.32}}{0.00}{\textbf{0.973}}{0.00} \\ 
\midrule
\rowcolor{surface}
\multicolumn{11}{c}{\bf Financial Certificate Tasks} \\
\midrule

\makecell[l]{CFA\\Level 1}
& \af{13.33}{0.00}{0.133}{0.00} 
& \af{42.22}{0.00}{0.418}{0.00} 
& \af{54.44}{0.00}{0.556}{0.00} 
& \af{63.33}{0.00}{0.631}{0.00} 
& \af{55.56}{0.00}{0.556}{0.00} 
& \af{86.67}{0.00}{0.867}{0.00} 
& \af{\textbf{87.78}}{0.00}{\textbf{0.878}}{0.00}
& \af{\textbf{87.78}}{0.00}{\textbf{0.878}}{0.00} 
& \af{\textbf{87.78}}{0.00}{\textbf{0.878}}{0.00} 
& \af{52.22}{0.00}{0.530}{0.00} \\ 

\makecell[l]{CFA\\Level 2}
& \af{19.48}{0.00}{0.199}{0.00} 
& \af{29.87}{0.00}{0.303}{0.00} 
& \af{46.75}{0.00}{0.485}{0.00} 
& \af{55.84}{0.00}{0.563}{0.00} 
& \af{56.67}{0.00}{0.567}{0.00} 
& \af{88.31}{0.00}{0.883}{0.00}
& \af{83.12}{0.00}{0.835}{0.00} 
& \af{90.91}{0.00}{0.909}{0.00} 
& \af{\textbf{92.21}}{0.00}{\textbf{0.922}}{0.00} 
& \af{51.11}{0.00}{0.519}{0.00} \\ 

\makecell[l]{CFA\\Level 3}
& \af{16.67}{0.00}{0.179}{0.00} 
& \af{24.36}{0.00}{0.271}{0.00} 
& \af{47.44}{0.00}{0.496}{0.00} 
& \af{51.28}{0.00}{0.517}{0.00} 
& \af{52.56}{0.00}{0.538}{0.00} 
& \af{70.51}{0.00}{0.705}{0.00}
& \af{66.67}{0.00}{0.675}{0.00} 
& \af{69.23}{0.00}{0.697}{0.00} 
& \af{\textbf{79.49}}{0.00}{\textbf{0.795}}{0.00} 
& \af{51.28}{0.00}{0.557}{0.00} \\ \addlinespace

\makecell[l]{CPA\\REG}
& \af{31.68}{0.00}{0.317}{0.00} 
& \af{41.58}{0.00}{0.426}{0.00} 
& \af{65.35}{0.00}{0.654}{0.00} 
& \af{67.33}{0.00}{0.667}{0.00} 
& \af{63.37}{0.00}{0.638}{0.00} 
& \af{80.20}{0.00}{0.802}{0.00} 
& \af{88.12}{0.00}{0.885}{0.00}
& \af{\textbf{90.10}}{0.00}{\textbf{0.901}}{0.00} 
& \af{\textbf{90.10}}{0.00}{\textbf{0.901}}{0.00} 
& \af{51.28}{0.00}{0.557}{0.00} \\  

\midrule

\rowcolor{surface}
\multicolumn{11}{c}{\bf Financial Reporting Tasks} \\
\midrule
FiNER 
& \af{21.28}{0.00}{0.232}{0.00} 
& \af{61.82}{0.00}{0.606}{0.00} 
& \af{68.92}{0.00}{0.699}{0.00} 
& \af{72.29}{0.00}{0.725}{0.00} 
& \af{63.91}{0.00}{0.638}{0.00} 
& \af{74.10}{0.00}{0.759}{0.00} 
& \af{74.32}{0.00}{0.760}{0.00}
& \af{70.92}{0.00}{0.732}{0.00} 
& \af{70.72}{0.00}{0.724}{0.00} 
& \af{\textbf{80.32}}{0.00}{\textbf{0.802}}{0.00} \\ 

FNXL 
& \af{3.64}{0.00}{0.045}{0.00} 
& \af{20.14}{0.00}{0.210}{0.00} 
& \af{27.33}{0.00}{0.288}{0.00} 
& \af{42.41}{0.00}{0.398}{0.00} 
& \af{37.75}{0.00}{0.356}{0.00} 
& \af{23.57}{0.00}{0.250}{0.00} 
& \af{23.05}{0.00}{0.253}{0.00}
& \af{33.50}{0.00}{0.311}{0.00} 
& \af{35.68}{0.00}{0.348}{0.00} 
& \af{\textbf{47.98}}{0.00}{\textbf{0.438}}{0.00} \\ \addlinespace

\makecell[l]{XBRL\\Term}
& \bert{-}{0.00}{0.574}{0.00} 
& \bert{-}{0.00}{0.587}{0.00} 
& \bert{-}{0.00}{0.573}{0.00} 
& \bert{-}{0.00}{0.584}{0.00} 
& \bert{-}{0.00}{0.572}{0.00} 
& \bert{-}{0.00}{0.599}{0.00}
& \bert{-}{0.00}{0.606}{0.00} 
& \bert{-}{0.00}{0.606}{0.00} 
& \bert{-}{0.00}{0.630}{0.00} 
& \bert{-}{0.00}{\textbf{0.666}}{0.00} \\ 
\midrule

\rowcolor{surface}
\multicolumn{11}{c}{\bf Financial Statement Analysis Tasks} \\
\midrule
\makecell[l]{Tag \\ Extraction}
& \af{69.16}{0.00}{0.739}{0.00} 
& \af{69.64}{0.00}{0.782}{0.00} 
& \af{85.03}{0.00}{0.849}{0.00} 
& \af{81.60}{0.00}{0.864}{0.00} 
& \af{80.27}{0.00}{0.811}{0.00} 
& \af{\textbf{89.13}}{0.00}{0.886}{0.00}
& \af{86.89}{0.00}{0.872}{0.00} 
& \af{80.44}{0.00}{0.896}{0.00} 
& \af{85.26}{0.00}{0.879}{0.00} 
& \af{85.03}{0.00}{\textbf{0.907}}{0.00} \\ 

\makecell[l]{Value \\ Extraction}
& \af{52.46}{0.00}{0.565}{0.00} 
& \af{88.19}{0.00}{0.904}{0.00} 
& \af{98.01}{0.00}{0.982}{0.00} 
& \af{97.01}{0.00}{0.974}{0.00} 
& \af{98.02}{0.00}{0.980}{0.00} 
& \af{98.49}{0.00}{0.986}{0.00} 
& \af{97.14}{0.00}{0.974}{0.00}
& \af{98.57}{0.00}{0.988}{0.00} 
& \af{99.13}{0.00}{\textbf{0.992}}{0.00} 
& \af{\textbf{99.20}}{0.00}{\textbf{0.992}}{0.00} \\ 

\makecell[l]{Formula \\ Construction}
& \af{12.92}{0.00}{0.201}{0.00} 
& \af{59.28}{0.00}{0.665}{0.00} 
& \af{22.75}{0.00}{0.315}{0.00} 
& \af{79.76}{0.00}{0.820}{0.00} 
& \af{61.90}{0.00}{0.644}{0.00} 
& \af{77.61}{0.00}{0.876}{0.00} 
& \af{89.34}{0.00}{\textbf{0.898}}{0.00}
& \af{88.02}{0.00}{0.882}{0.00} 
& \af{\textbf{89.46}}{0.00}{0.893}{0.00} 
& \af{67.85}{0.00}{0.786}{0.00} \\ 

\makecell[l]{Formula\\Calculation} 
& \af{27.27}{0.00}{0.317}{0.00} 
& \af{77.49}{0.00}{0.783}{0.00} 
& \af{85.99}{0.00}{0.868}{0.00} 
& \af{83.59}{0.00}{0.857}{0.00} 
& \af{53.57}{0.00}{0.536}{0.00} 
& \af{98.68}{0.00}{0.990}{0.00}
& \af{92.81}{0.00}{0.947}{0.00} 
& \af{\textbf{98.92}}{0.00}{\textbf{0.993}}{0.00} 
& \af{98.80}{0.00}{0.988}{0.00} 
& \af{54.76}{0.00}{0.548}{0.00} \\ \addlinespace

\makecell[l]{Finance\\Bench} 
& \bert{-}{0.00}{0.443}{0.00} 
& \bert{-}{0.00}{0.528}{0.00} 
& \bert{-}{0.00}{0.573}{0.00} 
& \bert{-}{0.00}{0.564}{0.00} 
& \bert{-}{0.00}{0.552}{0.00} 
& \bert{-}{0.00}{0.511}{0.00} 
& \bert{-}{0.00}{0.542}{0.00}
& \bert{-}{0.00}{0.477}{0.00} 
& \bert{-}{0.00}{\textbf{0.575}}{0.00} 
& \bert{-}{0.00}{0.544}{0.00} \\ \addlinespace

\makecell[l]{Financial\\Math} 
& \af{11.00}{0.00}{0.136}{0.00} 
& \af{10.50}{0.00}{0.134}{0.00} 
& \af{21.50}{0.00}{0.255}{0.00} 
& \af{27.00}{0.00}{0.296}{0.00} 
& \af{19.00}{0.00}{0.204}{0.00} 
& \af{30.00}{0.00}{0.332}{0.00}
& \af{26.50}{0.00}{0.307}{0.00} 
& \af{28.50}{0.00}{0.317}{0.00} 
& \af{34.50}{0.00}{0.370}{0.00} 
& \af{\textbf{66.00}}{0.00}{\textbf{0.785}}{0.00} \\

\midrule\rowcolor{surface}
\multicolumn{11}{c}{\textbf{Overall Average} (Using BERTScore F1 $\times$ 100)} \\

\midrule

\makecell[l]{Aggregated} 
& \textcolor{blue}{37.05}
& \textcolor{blue}{52.36}
& \textcolor{blue}{57.16}
& \textcolor{blue}{63.39}
& \textcolor{blue}{58.97}
& \textcolor{blue}{\textbf{74.74}}
& \textcolor{blue}{74.29}
& \textcolor{blue}{69.53}
& \textcolor{blue}{73.82}
& \textcolor{blue}{71.08} \\
\bottomrule

\end{tabular}
\label{tab:main_results_consolidated}

\end{table}

\begin{minipage}[t]{0.52\textwidth}
    \centering
    \small
    \captionof{table}{Comparison of fine-tuning cost. GPT-4o cost is estimated based on 4 epochs of fine-tuning at OpenAI fine-tuning pricing \cite{OpenAI_model_pricing}.} \label{table:finetuning_cost}
    \label{cost}
    \begin{tabular}{l|c|c|c}
    \toprule
    Models & \shortstack{ Time } & GPUs & \shortstack{Est. Cost \\ (USD)} \\
    \midrule
    BloombergGPT \cite{wu2023bloomberggpt} & 53 days & 512$\times$A100 & \$2.7 M \\
    \midrule
    \rowcolor{surface}
    LoRA   & 14.9h & 4 $\times$ A5000 & \$15.50 \\ 
    \rowcolor{surface}
    QLoRA  & 14.1h & 4 $\times$ A5000 & \$14.66 \\ 
    \rowcolor{surface}
    DoRA   & 15.9h & 4 $\times$ A5000 & \$16.54 \\ 
    \rowcolor{surface}
    rsLoRA & 14.5h & 4 $\times$ A5000 & \$15.11 \\ 
    \midrule
    Gemini 2.0 FL & 8.8h & - & \$162.02  \\
    GPT-4o-mini & - & - & \$312.00  \\
    \bottomrule
    \end{tabular}
\end{minipage} \hspace{12pt} 
\begin{minipage}[t]{0.44\textwidth} 
    \centering
    \captionof{figure}{Average inference time of LoRA fine-tuned Llama 3.1 8B and LoRA fine-tuned Gemini 2.0 FL across tasks}
    \label{figure:inference_time}
    \includegraphics[width=6cm]{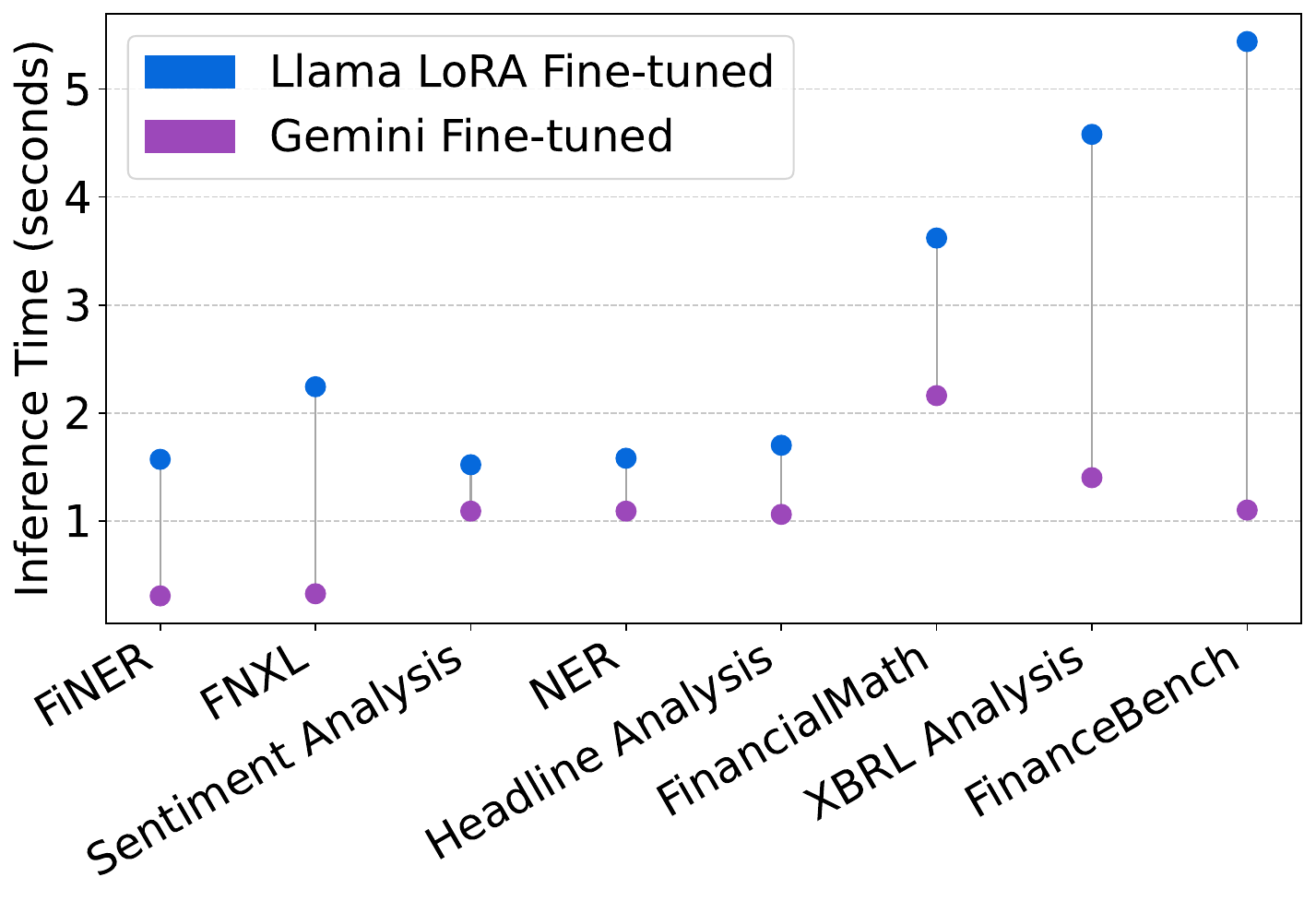} 
\end{minipage}

\newcommand{\vbase}[1]{\textcolor{black}{#1}}
\newcommand{\vsame}[1]{\textcolor{gray}{#1}}
\newcommand{\vup}[1]{\textcolor{green}{#1}} 
\newcommand{\vdown}[1]{\textcolor{error}{#1}}
\setlength{\tabcolsep}{2pt}

\begin{table*}[h]

  \begin{minipage}[t]{0.52\textwidth}
    \centering
    \small
    \caption{Accuracy on MMLU \& GSM8K benchmarks for Llama 3.1 8B base and eight LoRA adapters. Scores are colored relative to base: \vsame{gray} (same), \vup{green} (higher), \vdown{red} (lower)} 
    \label{tab:forgetting_mmlu_gsm8k_narrow}
    \begin{tabular}{l|c|cccc}
    \toprule
    \multirow{2}{*}{\textbf{Dataset}} & 
    \multirow{2}{*}{ \shortstack{\textbf{Llama}\\\textbf{3.1 8B}\\\textbf{(base)}} } &
    \multicolumn{4}{c}{\makecell[c]{\textbf{Llama 3.1 8B Adapters}}} \\
    \cmidrule(lr){3-6} 

    & &
    \makecell[c]{LoRA\\8bit‑r8} &
    \makecell[c]{QLoRA\\4bit‑r4} &
    \makecell[c]{DoRA\\8bit‑r8} &
    \makecell[c]{rsLoRA\\8bit‑r8} \\
    \midrule

    \makecell[l]{\textbf{MMLU} \cite{hendrycks2020measuring}\\(\textit{Sentiment})} &
    \vbase{0.229} & 
    \vsame{0.229} & 
    \vsame{0.229} & 
    \vsame{0.229} & 
    \vsame{0.229} \\ 

    \makecell[l]{\textbf{MMLU}\\(\textit{FiNER})} &
    \vbase{0.229} & 
    \vsame{0.229} & 
    \vsame{0.229} & 
    \vsame{0.229} & 
    \vsame{0.229} \\ 
    \midrule

    \makecell[l]{\textbf{GSM8K} \cite{cobbe2021training}\\(\textit{Sentiment})} &
    \vbase{0.011} & 
    \vup{0.014} &   
    \vup{0.014} &   
    \vsame{0.011} & 
    \vsame{0.011} \\ 

    \makecell[l]{\textbf{GSM8K}\\(\textit{FiNER})} &
    \vbase{0.011} & 
    \vsame{0.011} & 
    \vup{0.014} &   
    \vsame{0.011} & 
    \vup{0.016} \\   
    \bottomrule
    \end{tabular}
  \end{minipage} \hspace{12pt}
  \begin{minipage}[t]{0.46\textwidth} 
\setlength{\tabcolsep}{3pt}
  
    \centering
    \small
    \caption{Performance comparison of central LoRA and LoRA federated learning using four nodes on sentiment analysis tasks: accuracy (\acc{blue}) and F1 score (\fscore{gray}).} \label{table:federated}
    \label{tab:federated_performance_side} 
    \begin{tabular}{l|cccc} 
    \toprule
    \multirow{2}{*}{\makecell[l]{\textbf{Llama 3.1} \\ \textbf{8B 8bit-r8}}} & \multirow{2}{*}{FPB} & \multirow{2}{*}{FiQA SA} & \multirow{2}{*}{TFNS} & \multirow{2}{*}{NWGI}  \\ 
    & & & & \\ 
    \midrule
    \addlinespace
    \multirow{2}{*}{Base}
    & \acc{68.73} & \acc{46.55} & \acc{69.97} & \acc{46.58}  \\
    & \fscore{0.677} & \fscore{0.557} & \fscore{0.683} & \fscore{0.412}  \\ \addlinespace
    \vspace{1pt}
    \multirow{2}{*}{\makecell[l]{Central \\ (LoRA)}}
    & \textbf{\acc{89.11}} & \textbf{\acc{88.09}} & \textbf{\acc{91.96}} & \textbf{\acc{61.92}}  \\
    & \textbf{\fscore{0.941}} & \textbf{\fscore{0.923}} & \textbf{\fscore{0.955}} & \textbf{\fscore{0.748}} \\ [\smallskipamount]

    \midrule

    \multirow{2}{*}{FedAvg \cite{fedavg}} 
    & \acc{82.43} & \acc{76.17} & \acc{73.41} & \acc{56.02}  \\ 
    & \fscore{0.902} & \fscore{0.860} & \fscore{0.842} & \fscore{0.698}    \\

    \bottomrule
    \end{tabular}
  \end{minipage}
\end{table*}

\subsection{Angle III: Resource Usage and Performance Trade-offs of LoRA methods} \label{section:resource}

Table~\ref{table:finetuning_cost} details the computational costs of LoRA fine-tuned models. Using four NVIDIA A5000 GPUs, the wall-clock time for fine-tuning ranged from 14.1 hours (QLoRA) to 15.9 hours (DoRA), corresponding to a total of approximately 56.4 to 63.6 GPU hours. At an estimated rate of \$0.26 per GPU hour, this translates to a cost of roughly \$14.66 to \$16.54. This is substantially more cost-effective than fine-tuning services from providers like Google or OpenAI. Figure~\ref{figure:inference_time} illustrates the inference time of fine-tuned models on various datasets. Gemini API generally exhibits lower inference latency and is less sensitive to increasing prompt lengths than local Llama 3.1 8B Instruct inference, even when accounting for network overhead for the API. However, the inference speed of locally deployed Llama models can be significantly enhanced through the use of larger batch sizes.

\subsection{Angle IV: Practicability of Applying LoRA in Real-world Financial Scenarios}
\paragraph{Federated LoRA} 
The sensitive nature of financial data necessitates privacy-preserving techniques like Federated Learning for collaborative training. To explore this, we evaluated Federated LoRA~\cite{sun2024improvingloraprivacypreservingfederated}, with results presented in Table~\ref{table:federated}. Our experimental setup simulated a four-node environment employing the FedAvg algorithm~\cite{fedavg}, where the sentiment analysis dataset was partitioned across these nodes. The performance of this approach was benchmarked against both the base Llama model and standard centralized LoRA fine-tuning. While Federated LoRA did not match the performance levels of centralized LoRA, the results demonstrate a notable improvement compared to the base Llama model.
\paragraph{Catastrophic Forgetting} A major concern with PEFT is that fine-tuning on domain-specific tasks leads to the model forgetting pre-training knowledge. To investigate this, we evaluated eight adapters—covering both sentiment and FiNER tasks and all four LoRA variants—as well as the Llama 3.1 8B Instruct base model on two out-of-domain benchmarks, MMLU \cite{hendrycks2020measuring} and GSM8K \cite{cobbe2021training}. We used a zero-shot, no chain-of-thought setting to isolate stored knowledge. Table~\ref{tab:forgetting_mmlu_gsm8k_narrow} shows identical MMLU accuracy across all adapters and the base model, and equal or higher scores on GSM8K. Hence, at the ranks $r$ we tested (4 and 8) with $\alpha$:$r$ equal to 8:1 or 4:1, we observe that LoRA does not exhibit catastrophic forgetting. In fact, the slight GSM8K performance improvements hint at cross-domain knowledge transfer—fine-tuning on financial data may improve the model’s numerical reasoning skills.

\section{Related Works}
\label{section:related_work}

\subsection{Financial LLMs and Benchmarks}

BloombergGPT~\cite{wu2023bloomberggpt} is the first LLM specialized for the financial domain. The-50-billion parameter model was trained from scratch using a mix of financial and general datasets. The evaluation was conducted on a series of financial tasks including sentiment analysis, named entity recognition (NER), and question answering (QA) as well as general benchmarks, showing performance exceeding comparable models on financial tasks and strong performance on general tasks. 

FinGPT~\cite{liu2023data,liu2024differentially,icdcs} aims to provide a customized and personalized financial LLM. Instead of training from the ground up, FinGPT applied LoRA fine-tuning on open-source LLMs using general financial training sets. Performance evaluation displayed noticeable improvement over the base model, even surpassing that of BloombergGPT, while having substantial memory reduction and training speedup compared to training-from-scratch.

FinBen~\cite{xie2024finben} and PIXIU~\cite{xie2023pixiu} are financial benchmarks that offer a broad array of curated datasets. Benchmarking various general LLMs, they conclude that while LLMs demonstrate strong capabilities in textual analysis, they face challenges with advanced reasoning and complex financial problem-solving.

\subsection{Parameter-Efficient Fine-Tuning (PEFT) with Low-rank Adaptation (LoRA) Methods}

Full fine-tuning, which fine-tunes the full parameters of an LLM, is extremely computationally expensive. Parameter-efficient fine-tuning (PEFT) was proposed to reduce the number of trainable parameters by only fine-tuning a small number of model parameters~\cite{Mao_2024}. Low-rank adaptation (LoRA)~\cite{lora} is a widely used PEFT method that inserts a smaller set of pluggable low-rank trainable weights. The performance of downstream tasks after LoRA fine-tuning is comparable to that of full fine-tuning. Quantized LoRA (QLoRA)~\cite{qlora} quantizes the LLM to 4 bits and applies LoRA fine-tuning on such a model. QLoRA can significantly reduce GPU memory usage. 

\subsection{LoRA Methods with Federated Learning}

In the financial domain, private training data might be spread across multiple institutions. To fine-tune LLMs with non-centralized data, federated learning is needed. Several research papers have applied LoRA on federated learning, such as PrivateLoRA~\cite{wang2023privateloraefficientprivacypreserving} and Federated Freeze A LoRA (FFA-LoRA)~\cite{sun2024improvingloraprivacypreservingfederated}. 




\section{Conclusion and Future Work}

In this paper, we present FinLoRA, a benchmark that evaluates LoRA methods on both general and highly specialized financial tasks. We curated 19 diverse datasets covering a wide range of financial applications. Our study includes 46 rounds of fine-tuning and 194 rounds of evaluation to thoroughly assess and analyze commonly used LoRA methods. FinLoRA offers insights into overall performance, task-specific results, resource requirements for fine-tuning and inference, and practical considerations for real-world deployment—including data privacy in collaborative training and catastrophic forgetting. Our results demonstrate that fine-tuning can significantly enhance the effectiveness of LLMs on financial tasks. Additionally, FinLoRA provides a comprehensive collection of datasets with baseline results, laying a solid foundation for future research in this field. Moving forward, we plan to expand FinLoRA by incorporating additional LoRA methods into the project.



\bibliographystyle{plain}

\bibliography{neurips}

\begin{thebibliography}{10}

\bibitem{bhatia-etal-2024-fintral}
Gagan Bhatia, El~Moatez~Billah Nagoudi, Hasan Cavusoglu, and Muhammad Abdul-Mageed.
\newblock {F}in{T}ral: A family of {GPT}-4 level multimodal financial large language models.
\newblock In Lun-Wei Ku, Andre Martins, and Vivek Srikumar, editors, {\em Findings of the Association for Computational Linguistics: ACL 2024}, pages 13064--13087, August 2024.

\bibitem{callanan-etal-2024-gpt}
Ethan Callanan, Amarachi Mbakwe, Antony Papadimitriou, Yulong Pei, Mathieu Sibue, Xiaodan Zhu, Zhiqiang Ma, Xiaomo Liu, and Sameena Shah.
\newblock Can {GPT} models be financial analysts? an evaluation of {C}hat{GPT} and {GPT}-4 on mock {CFA} exams.
\newblock In {\em Proceedings of the Eighth Financial Technology and Natural Language Processing and the 1st Agent AI for Scenario Planning}, pages 23--32, Jeju, South Korea, 3 August 2024. -.

\bibitem{chen2021data}
Lulu Chen, Yingzhou Lu, Chiung-Ting Wu, Robert Clarke, Guoqiang Yu, Jennifer~E Van~Eyk, David~M Herrington, and Yue Wang.
\newblock Data-driven detection of subtype-specific differentially expressed genes.
\newblock {\em Scientific Reports}, 11(1):332, 2021.

\bibitem{chen2022adaptformer}
Shoufa Chen, Chongjian Ge, Zhan Tong, Jiangliu Wang, Yibing Song, Jue Wang, and Ping Luo.
\newblock Adaptformer: Adapting vision transformers for scalable visual recognition.
\newblock {\em Advances in Neural Information Processing Systems}, 35:16664--16678, 2022.

\bibitem{lu2024uncertainty}
Tianyi Chen, Nan Hao, Capucine Van~Rechem, Jintai Chen, and Tianfan Fu.
\newblock Uncertainty quantification and interpretability for clinical trial approval prediction.
\newblock {\em Health Data Science}, 4:0126, 2024.

\bibitem{cobbe2021training}
Karl Cobbe, Vineet Kosaraju, Mohammad Bavarian, Mark Chen, Heewoo Jun, Lukasz Kaiser, Matthias Plappert, Jerry Tworek, Jacob Hilton, Reiichiro Nakano, et~al.
\newblock Training verifiers to solve math word problems.
\newblock {\em arXiv preprint arXiv:2110.14168}, 2021.

\bibitem{qlora}
Tim Dettmers, Artidoro Pagnoni, Ari Holtzman, and Luke Zettlemoyer.
\newblock {QL}o{RA}: Efficient finetuning of quantized {LLM}s.
\newblock In {\em Thirty-seventh Conference on Neural Information Processing Systems}, 2023.

\bibitem{llama}
Abhimanyu Dubey, Abhinav Jauhri, Abhinav Pandey, Abhishek Kadian, Ahmad Al-Dahle, Aiesha Letman, Akhil Mathur, and et~al.
\newblock The {Llama} 3 herd of models, 2024.

\bibitem{gao2020pile}
Leo Gao, Stella Biderman, Sid Black, Laurence Golding, Travis Hoppe, Charles Foster, Jason Phang, Horace He, Anish Thite, Noa Nabeshima, et~al.
\newblock The {Pile}: An {800GB} dataset of diverse text for language modeling.
\newblock {\em arXiv preprint arXiv:2101.00027}, 2020.

\bibitem{xbrlagent}
Shijie Han, Haoqiang Kang, Bo~Jin, Xiao-Yang Liu, and Steve~Y Yang.
\newblock {XBRL Agent}: Leveraging large language models for financial report analysis.
\newblock In {\em Proceedings of the 5th ACM International Conference on AI in Finance}, ICAIF '24, page 856–864, 2024.

\bibitem{hendrycks2020measuring}
Dan Hendrycks, Collin Burns, Steven Basart, Andy Zou, Mantas Mazeika, Dawn Song, and Jacob Steinhardt.
\newblock Measuring massive multitask language understanding.
\newblock {\em arXiv preprint arXiv:2009.03300}, 2020.

\bibitem{lora}
Edward~J Hu, Yelong shen, Phillip Wallis, Zeyuan Allen-Zhu, Yuanzhi Li, Shean Wang, Lu~Wang, and Weizhu Chen.
\newblock Lo{RA}: Low-rank adaptation of large language models.
\newblock In {\em International Conference on Learning Representations}, 2022.

\bibitem{huanlora}
Muchen Huan and Jianhong Shun.
\newblock Fine-tuning transformers efficiently: A survey on {LoRA} and its impact.
\newblock {\em Preprints}, February 2025.

\bibitem{hurst2024gpt}
Aaron Hurst, Adam Lerer, Adam~P Goucher, Adam Perelman, Aditya Ramesh, Aidan Clark, AJ~Ostrow, Akila Welihinda, Alan Hayes, Alec Radford, et~al.
\newblock {GPT}-4o system card.
\newblock {\em arXiv preprint arXiv:2410.21276}, 2024.

\bibitem{islam2023financebenchnewbenchmarkfinancial}
Pranab Islam, Anand Kannappan, Douwe Kiela, Rebecca Qian, Nino Scherrer, and Bertie Vidgen.
\newblock {FinanceBench}: A new benchmark for financial question answering, 2023.

\bibitem{kalajdzievski2023rankstabilizationscalingfactor}
Damjan Kalajdzievski.
\newblock {A Rank Stabilization Scaling Factor for Fine-Tuning with LoRA}, 2023.

\bibitem{Lee_2025}
Jean Lee, Nicholas Stevens, and Soyeon~Caren Han.
\newblock Large language models in finance (finllms).
\newblock {\em Neural Computing and Applications}, January 2025.

\bibitem{liu2024deepseek}
Aixin Liu, Bei Feng, Bing Xue, Bingxuan Wang, Bochao Wu, Chengda Lu, Chenggang Zhao, Chengqi Deng, Chenyu Zhang, Chong Ruan, et~al.
\newblock {DeepSeek-V3} technical report.
\newblock {\em arXiv preprint arXiv:2412.19437}, 2024.

\bibitem{liu2022few}
Haokun Liu, Derek Tam, Mohammed Muqeeth, Jay Mohta, Tenghao Huang, Mohit Bansal, and Colin~A Raffel.
\newblock Few-shot parameter-efficient fine-tuning is better and cheaper than in-context learning.
\newblock {\em Advances in Neural Information Processing Systems}, 35:1950--1965, 2022.

\bibitem{liu2024socraticlm}
Jiayu Liu, Zhenya Huang, Tong Xiao, Jing Sha, Jinze Wu, Qi~Liu, Shijin Wang, and Enhong Chen.
\newblock Socratic{LM}: Exploring socratic personalized teaching with large language models.
\newblock In {\em The Thirty-eighth Annual Conference on Neural Information Processing Systems}, 2024.

\bibitem{liu2024dora}
Shih-Yang Liu, Chien-Yi Wang, Hongxu Yin, Pavlo Molchanov, Yu-Chiang~Frank Wang, Kwang-Ting Cheng, and Min-Hung Chen.
\newblock {DoRA}: Weight-decomposed low-rank adaptation.
\newblock In {\em Forty-first International Conference on Machine Learning}, 2024.

\bibitem{liu2023data}
Xiao-Yang Liu, Guoxuan Wang, Hongyang Yang, and Daochen Zha.
\newblock Data-centric {FinGPT}: Democratizing internet-scale data for financial large language models.
\newblock In {\em Workshop on Instruction Tuning and Instruction Following, NeurIPS}, 2023.

\bibitem{icdcs}
Xiao-Yang Liu, Jie Zhang, Guoxuan Wang, Weiqin Tong, and Anwar Walid.
\newblock {Efficient Pretraining and Finetuning of Quantized LLMs with Low-Rank Structure }.
\newblock In {\em IEEE 44th International Conference on Distributed Computing Systems (ICDCS)}, pages 300--311, July 2024.

\bibitem{liu2024differentially}
Xiao-Yang Liu, R.~Zhu, Daochen Zha, J.~Gao, S.~Zhong, Matt White, and Meikang Qiu.
\newblock Differentially private low-rank adaptation of large language model using federated learning.
\newblock {\em ACM Transactions on Management Information Systems}, 2024.

\bibitem{loukas-etal-2022-finer}
Lefteris Loukas, Manos Fergadiotis, Ilias Chalkidis, Eirini Spyropoulou, Prodromos Malakasiotis, Ion Androutsopoulos, and Georgios Paliouras.
\newblock {F}i{NER}: Financial numeric entity recognition for {XBRL} tagging.
\newblock In {\em Proceedings of the 60th Annual Meeting of the Association for Computational Linguistics (Volume 1: Long Papers)}, Dublin, Ireland, May 2022.

\bibitem{lu2022cot}
Yingzhou Lu, Chiung-Ting Wu, Sarah~J Parker, Zuolin Cheng, Georgia Saylor, Jennifer~E Van~Eyk, Guoqiang Yu, Robert Clarke, David~M Herrington, and Yue Wang.
\newblock {COT}: an efficient and accurate method for detecting marker genes among many subtypes.
\newblock {\em Bioinformatics Advances}, 2(1), 2022.

\bibitem{fiqa}
Macedo Maia, Siegfried Handschuh, Andre Freitas, Brian Davis, Ross McDermott, Manel Zarrouk, and Alexandra Balahur.
\newblock Www'18 open challenge: Financial opinion mining and question answering.
\newblock pages 1941--1942, 04 2018.

\bibitem{fpb}
Pekka Malo, Ankur Sinha, Pyry Takala, Pekka Korhonen, and Jyrki Wallenius.
\newblock Good debt or bad debt: Detecting semantic orientations in economic texts, 2013.

\bibitem{Mao_2024}
Yuren Mao, Yuhang Ge, Yijiang Fan, Wenyi Xu, Yu~Mi, Zhonghao Hu, and Yunjun Gao.
\newblock A survey on lora of large language models.
\newblock {\em Frontiers of Computer Science}, 19(7), December 2024.

\bibitem{fedavg}
Brendan McMahan, Eider Moore, Daniel Ramage, Seth Hampson, and Blaise Aguera~y Arcas.
\newblock {Communication-Efficient Learning of Deep Networks from Decentralized Data}.
\newblock In Aarti Singh and Jerry Zhu, editors, {\em Proceedings of the 20th International Conference on Artificial Intelligence and Statistics}, volume~54 of {\em Proceedings of Machine Learning Research}, pages 1273--1282. PMLR, 20--22 Apr 2017.

\bibitem{meng2024pissa}
Fanxu Meng, Zhaohui Wang, and Muhan Zhang.
\newblock Pissa: Principal singular values and singular vectors adaptation of large language models.
\newblock {\em Advances in Neural Information Processing Systems}, 37:121038--121072, 2024.

\bibitem{OpenAI_model_pricing}
{OpenAI}.
\newblock {OpenAI API} pricing.
\newblock \url{https://platform.openai.com/docs/pricing}, May 2025.
\newblock Accessed: 2025-05-14.

\bibitem{tfns}
Md.~Abdur Rahman.
\newblock Twitter financial news sentiment.
\newblock \url{http://precog.iiitd.edu.in/people/anupama}, 2022.

\bibitem{xbrl}
Ali Saeedi, Jim Richards, and Barry Smith.
\newblock An introduction to {XBRL}.
\newblock In {\em British Accounting Association{\textquoteright}s Annual Conference}, 2007.

\bibitem{ner}
Julio~Cesar Salinas~Alvarado, Karin Verspoor, and Timothy Baldwin.
\newblock Domain adaption of named entity recognition to support credit risk assessment.
\newblock In Ben Hachey and Kellie Webster, editors, {\em Proceedings of the Australasian Language Technology Association Workshop 2015}, pages 84--90, Parramatta, Australia, December 2015.

\bibitem{fnxl}
Soumya Sharma, Subhendu Khatuya, Manjunath Hegde, Afreen Shaikh, Koustuv Dasgupta, Pawan Goyal, and Niloy Ganguly.
\newblock Financial numeric extreme labelling: A dataset and benchmarking.
\newblock In {\em Findings of the Association for Computational Linguistics: ACL 2023}, pages 3550--3561, July 2023.

\bibitem{headline}
Ankur Sinha and Tanmay Khandait.
\newblock Impact of news on the commodity market: Dataset and results, 2020.

\bibitem{sun2024improvingloraprivacypreservingfederated}
Youbang Sun, Zitao Li, Yaliang Li, and Bolin Ding.
\newblock Improving lo{RA} in privacy-preserving federated learning.
\newblock In {\em The Twelfth International Conference on Learning Representations}, 2024.

\bibitem{geminiteam2024geminifamilyhighlycapable}
Gemini Team, Rohan Anil, Sebastian Borgeaud, et~al.
\newblock Gemini: A family of highly capable multimodal models, 2024.

\bibitem{wang2023privateloraefficientprivacypreserving}
Yiming Wang, Yu~Lin, Xiaodong Zeng, and Guannan Zhang.
\newblock {PrivateLoRA} for efficient privacy preserving {LLM}, 2023.

\bibitem{wang2024twin}
Yue Wang, Tianfan Fu, Yinlong Xu, Zihan Ma, Hongxia Xu, Bang Du, Yingzhou Lu, Honghao Gao, Jian Wu, and Jintai Chen.
\newblock {TWIN-GPT}: Digital twins for clinical trials via large language model.
\newblock {\em ACM Trans. Multimedia Comput. Commun. Appl.}, July 2024.
\newblock Just Accepted.

\bibitem{wu2023bloomberggpt}
Shijie Wu, Ozan Irsoy, Steven Lu, Vadim Dabravolski, Mark Dredze, Sebastian Gehrmann, Prabhanjan Kambadur, David Rosenberg, and Gideon Mann.
\newblock {BloombergGPT}: A large language model for finance.
\newblock {\em arXiv preprint arXiv:2303.17564}, 2023.

\bibitem{DBLP:conf/acl/WuWGL24}
Yang Wu, Chenghao Wang, Ece Gumusel, and Xiaozhong Liu.
\newblock Knowledge-infused legal wisdom: Navigating llm consultation through the lens of diagnostics and positive-unlabeled reinforcement learning.
\newblock In {\em ACL (Findings)}, pages 15542--15555, 2024.

\bibitem{xie2024finben}
Qianqian Xie, Weiguang Han, Zhengyu Chen, Ruoyu Xiang, Xiao Zhang, Yueru He, Mengxi Xiao, Dong Li, Yongfu Dai, Duanyu Feng, Yijing Xu, Haoqiang Kang, Ziyan Kuang, Chenhan Yuan, Kailai Yang, Zheheng Luo, Tianlin Zhang, Zhiwei Liu, Guojun Xiong, Zhiyang Deng, Yuechen Jiang, Zhiyuan Yao, Haohang Li, Yangyang Yu, Gang Hu, Huang Jiajia, Xiao-Yang Liu, Alejandro Lopez-Lira, Benyou Wang, Yanzhao Lai, Hao Wang, Min Peng, Sophia Ananiadou, and Jimin Huang.
\newblock {FinBen}: An holistic financial benchmark for large language models.
\newblock In {\em The Thirty-eight Conference on Neural Information Processing Systems Datasets and Benchmarks Track}, 2024.

\bibitem{xie2023pixiu}
Qianqian Xie, Weiguang Han, Xiao Zhang, Yanzhao Lai, Min Peng, Alejandro Lopez-Lira, and Jimin Huang.
\newblock {PIXIU}: A comprehensive benchmark, instruction dataset and large language model for finance.
\newblock In {\em Thirty-seventh Conference on Neural Information Processing Systems Datasets and Benchmarks Track}, 2023.

\bibitem{yang2024lowrankadaptationfoundationmodels}
Menglin Yang, Jialin Chen, Yifei Zhang, Jiahong Liu, Jiasheng Zhang, Qiyao Ma, Harshit Verma, Qianru Zhang, Min Zhou, Irwin King, and Rex Ying.
\newblock Low-rank adaptation for foundation models: A comprehensive review, 2024.

\bibitem{zhang2023enhancingfinancialsentimentanalysis}
Boyu Zhang, Hongyang Yang, Tianyu Zhou, Muhammad Ali~Babar, and Xiao-Yang Liu.
\newblock Enhancing financial sentiment analysis via retrieval augmented large language models.
\newblock In {\em ACM International Conference on AI in Finance}, pages 349--356, 2023.

\bibitem{zhang2020bertscoreevaluatingtextgeneration}
Tianyi Zhang, Varsha Kishore, Felix Wu, Kilian~Q Weinberger, and Yoav Artzi.
\newblock Bertscore: Evaluating text generation with bert.
\newblock In {\em International Conference on Learning Representations}, 2020.

\bibitem{zhao2023survey}
Wayne~Xin Zhao, Kun Zhou, Junyi Li, Tianyi Tang, Xiaolei Wang, Yupeng Hou, Yingqian Min, Beichen Zhang, Junjie Zhang, Zican Dong, et~al.
\newblock A survey of large language models.
\newblock {\em arXiv preprint arXiv:2303.18223}, 1(2), 2023.

\bibitem{zheng2023judging}
Lianmin Zheng, Wei-Lin Chiang, Ying Sheng, Siyuan Zhuang, Zhanghao Wu, Yonghao Zhuang, Zi~Lin, Zhuohan Li, Dacheng Li, Eric Xing, et~al.
\newblock Judging llm-as-a-judge with mt-bench and chatbot arena.
\newblock {\em Advances in Neural Information Processing Systems}, 36:46595--46623, 2023.

\bibitem{zhou2024large}
Shuang Zhou, Zidu Xu, Mian Zhang, Chunpu Xu, Yawen Guo, Zaifu Zhan, Sirui Ding, Jiashuo Wang, Kaishuai Xu, Yi~Fang, et~al.
\newblock Large language models for disease diagnosis: A scoping review.
\newblock {\em arXiv preprint arXiv:2409.00097}, 2024.

\end{thebibliography}

\end{document}